\documentclass[journal]{IEEEtran}
\usepackage{cite}
\usepackage{amsmath,amssymb,amsfonts}
\usepackage{algorithm}
\usepackage{algpseudocode}
\usepackage{graphicx}
\usepackage{textcomp}
\usepackage[export]{adjustbox}
\usepackage{xcolor}
\usepackage{gensymb}
\usepackage{listings}
\usepackage{url}
\usepackage[margins]{trackchanges}

\usepackage{footnote}  

\usepackage{ulem}    

\ifCLASSINFOpdf
\else
\fi
\begin{document}
%
\title{Optical Linearization of Silicon Photonic Ring-Assisted Mach-Zehnder Modulator}
%
%
%

\author{Md~Jubayer Shawon,~\IEEEmembership{Student Member,~IEEE,}
        and~Vishal~Saxena,~\IEEEmembership{Senior Member,~IEEE}
\thanks{The authors are with the Department of Electrical and Computer Engineering, University of Delaware, Newark,
DE, 19716 USA e-mail: shawon@udel.edu.}
}

\maketitle
\thispagestyle{empty}
\pagestyle{empty}

\begin{abstract}
In high-performance RF photonic systems, the Electro-Optic (EO) modulators play a critical role as a key component, requiring low SWaP-C and high linearity. While traditional lithium niobate (LiNbO$_3$) Mach-Zehnder Modulators (MZMs) have been extensively utilized due to their superior linearity, silicon-based EO modulators have lagged behind in achieving comparable performance. This paper presents an experimental demonstration of a Ring Assisted Mach Zehnder Modulator (RAMZM) fabricated using a silicon photonic foundry process, addressing the performance gap. The proposed RAMZM modulator enables linearization in the optical domain and can be dynamically reconfigured to linearize around user-specified center frequency and bias conditions, even in the presence of process variations and thermal crosstalk. An automatic reconfiguration algorithm, empowered by Digital-to-Analog Converters (DACs), Analog-to-Digital Converters (ADCs), Trans-Impedance Amplifiers (TIAs), and a digital configuration engine, is developed to achieve linearization, resulting in a spurious-free dynamic range (SFDR) exceeding 113 dB.Hz$^{2/3}$. Furthermore, a biasing scheme is introduced for RAMZMs, significantly enhancing the modulation slope efficiency, which in turn yields a tone gain of over 13 dB compared to its standard operation. This reconfigurable electro-optic modulator can be seamlessly integrated into integrated RF photonic System-on-Chips (SoCs), leveraging the advantages of integration and cost-effectiveness.
\end{abstract}

\begin{IEEEkeywords}
Analog optical link, Silicon Photonics, Photonic Integrated Circuit (PIC), Ring-Assisted Mach Zehnder Modulator (RAMZM), RF Photonics, RF-to-optical modulator.
\end{IEEEkeywords}

%
\IEEEpeerreviewmaketitle

\section{Introduction}
%
%
%
%
\IEEEPARstart{E}{lectro}-Optic (EO) modulators are one of the key elements of any RF photonic system. Traditionally, discrete lithium niobate (LiNbO$_3$) MZMs have widely been used in high-performance RF photonic systems where linearity is of great importance. However, as system complexity grows, there is a significant demand for EO modulators realized in a CMOS-compatible platform, primarily due to the effortless co-integration of the modulator with control electronics and other RF System-on-Chip (SoC) components in such platform. Silicon photonics has recently emerged as a disruptive technology platform where RF photonic components can be fabricated alongside electronics, leveraging CMOS manufacturing ecosystem. Nonetheless, silicon photonic electro-optic modulators have not achieved the same level of performance as their matured LiNbO$_3$ counterparts, particularly when highly linear operation is required. Thus, achieving linearization of silicon-based electro-optic modulators has become a compelling area of research.

Various approaches have been explored to linearize silicon-based MZMs, including the utilization of electronic pre-distortion circuits \cite{okyere2017fifth} and third-order intermodulation (IM3) cancellation techniques \cite{hosseinzadeh2020distributed} in the electronic domain. However, these methods are inherently narrowband approaches, failing to fully exploit the broadband capabilities offered by photonics. On the other hand, Ring-Assisted MZ Modulators (RAMZM) have shown great promise by harnessing the expansive voltage-to-phase response of the ring to linearize the compressive phase-to-amplitude response of MZMs \cite{cardenas2013linearized}. However, existing RAMZM implementations have relied on manual tuning \cite{zhang2016ultralinear}, limiting their broader adoption in large-scale RF photonic systems. Therefore, in this paper, we present a novel silicon-based reconfigurable RAMZM along with an automatic tuning algorithm, aiming to achieve highly linear operation.

\section{RF Photonic Links}

\begin{figure}[!t]
\centering
\includegraphics[width=1\columnwidth]{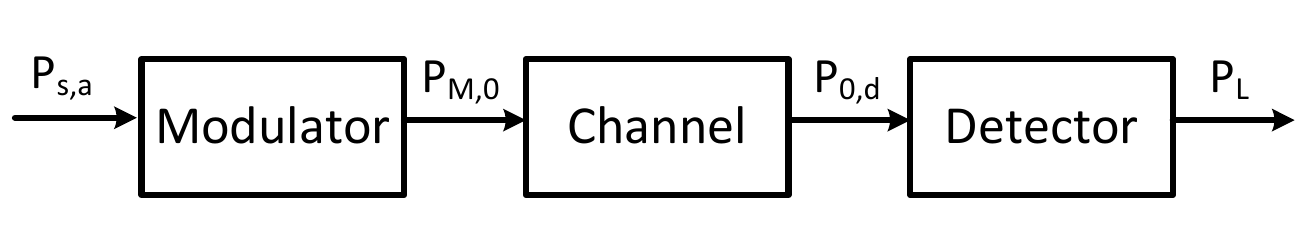}
\caption{An RF Photonic link using EO modulator. Here, $P_{M,o}$ is the output optical power of the modulator, and $P_{o,d}$ is the optical input power to the detector. $P_{s,a}$ and $P_L$ are the available input and output powers of the link.}
\label{fig:analog_link} 
\end{figure}

As shown in Fig. \ref{fig:analog_link}, an analog RF photonic optical link employs either a direct modulated laser (DML) or a Mach-Zehnder Modulator (MZM) which modulates a continuous-wave (CW) laser source (typically at the 1550nm telecommunication wavelength)  using an RF signal. The RF signal in the optical domain undergoes optical signal processing (e.g. filtering, phased-array processing, etc.) and transmitted over optical fiber(s). The optical-domain RF signal is finally converted back into RF domain using a (photo)detector, which is optionally followed by a low-noise transimpedance amplifier (LNTIA). 
In Fig. \ref{fig:analog_link}, $P_{s,a}$ is the input RF source power, $P_{M,o}$ is the optical power at the modulator output, $P_{o,d}$ is the optical power at the detector input, $P_{L}$ is the RF power at the load. The link gain ($G$), determined as
\begin{equation} \label{eq:LinkGain}
	G = G_{M} \cdot G_{ch}^2 \cdot G_{det}
\end{equation}
where $G_{M} \triangleq \frac{P_{M,o}^2}{P_{s,a}}$, $G_{ch} \triangleq \frac{P_{o,d}}{P_{M,o}}$, $G_{det} \triangleq \frac{P_L}{P_{o,d}^2}$ are the incremental modulation efficiency, optical channel gain, and the incremental detector efficiency, respectively.

\section{Ring-Assisted Mach-Zehnder Modulator}

\begin{figure}[!tbh]
\centering
\includegraphics[width=1\columnwidth]{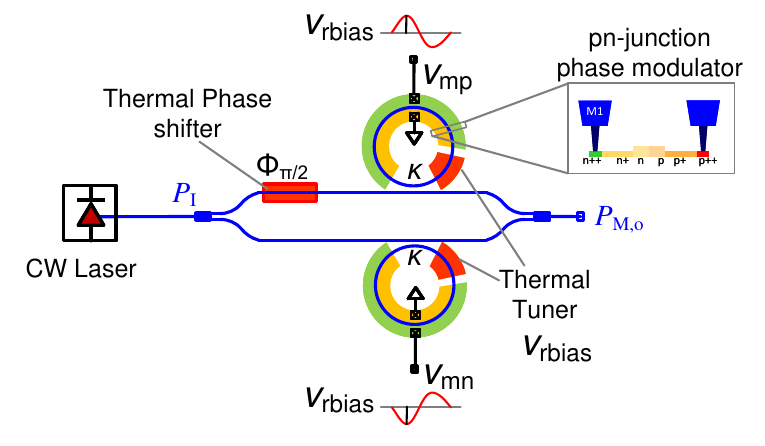}
\caption{Conceptual schematic of a silicon photonic Ring Assisted Mach-Zehnder Modulator (RAMZM), interfaced with RF and a CW laser source. }
\label{fig:ramzi}
\end{figure}

Ring-assisted Mach-Zehnder modulators (RAMZI) are similar to traditional Mach-Zehnder modulators (MZM)  but differ in that either one or both of the two phase shifters (upper and lower) are replaced by ring modulators, as shown in Fig. \ref{fig:ramzi}. By exploiting the Lorentzian transmission characteristics, upper and lower rings can be optically biased such that their voltage-to-phase responses are supra-linear. This supra-linear response can be enhanced by a suitable value of coupling coefficient ($\kappa$) and then used to compensate for the sub-linear phase-to-intensity non-linearity of a Mach-Zehnder interferometer \cite{gutierrez2012ring,cardenas2013linearized,morton2013morton,mwscas18a}. This approach effectively reduces the third-order non-linearity, while the parent MZM when biased at quadrature, suppresses the even-order non-linearity. Consequently, the output of the RAMZM can be made highly linear.

As illustrated in Fig. \ref{fig:ramzi}, the output of the continuous wave (CW) laser is equally divided between the two arms of the MZM. RF signals, applied to the ring modulators in a differential push-pull configuration around a DC bias ($V_{M}$), are translated into varying phase shifts in each arm of the MZI. When these two optical signals combine at the output of the MZM, optical modulation occurs via either constructive or destructive interference \cite{chrostowski2015silicon}.

\subsection{RAMZM Transmission Characteristics}
We now derive the transmission characteristics of the RAMZM. The electric field at the output of a RAMZM can be expressed as \cite{yue2013mmi}-

\begin{equation} \label{eq:ein}
\begin{aligned}
E_{out} &= \frac{E_{in}}{2} \Big( \left | a_{r1}(\theta) \right |e^{-j.(knL_1+ \phi_{quad} +\angle a_{r1}(\theta))} + \cdots \\ 
&+ \left | a_{r2}(\theta) \right |e^{-j.(knL_2+ \angle a_{r2}(\theta))} \Big)
\end{aligned}
\end{equation}

\begin{equation} \label{eq:artht}
\begin{aligned}
a_{ri}(\theta) = \frac{\tau - \alpha e^{-j\theta}}{1 - \tau \alpha e^{-j\theta}}
\end{aligned}
\end{equation}

where the variables are defined below. Here, index term $i=1$ refers to upper while $i=2$ refers to the lower arm of RAMZM. \\ 
$E_{in}$: input electric field\\
$L_i$: length of the RAMZM arms\\
$\left | a_{ri}(\theta)\right |$: ring magnitude response\\
$\angle a_{ri}(\theta)$: ring phase response\\
$\phi_{quad}$: quadrature bias of the Mach Zehnder\\ 
$\tau=\sqrt{1-\kappa^2}$: transmission coefficient of the ring couplers \\
$\alpha$: ring loss factor\\
$\theta=\theta(V)$: ring roundtrip phase shift, which depends on the applied voltage in the phase-modulators. \\

If we neglect losses in the short MZI arms, the transmission function, $T(\theta) = \frac{I_{out}}{I_{in}}= \Big|\frac{E_{out}}{E_{in}}\Big|^2$, can be written as


\begin{align} \label{eq:tf}
T(\theta) 
&= \frac{1}{4}\left|e^{-j.(knL_1+\phi_{quad} + \angle a_{r1}(\theta))} + e^{-j.(knL_2+ \angle a_{r2}(\theta))} \right|^2
\end{align}


When the optical path of the RAMZM arms have equal length ($L_2 = L_1$), the transfer function can be simplified as  

\begin{align}
T(\theta) &= \frac{1}{2}\left [ 1+\cos \Big(\phi_{quad}+\angle a_{r1}(\theta)-\angle a_{r2}(\theta) \Big) \right ] \label{eq:tf2} 
\end{align}

\begin{equation} \label{eq:aris}
\begin{aligned}
\phi_i = \angle a_{ri}(\theta) = \tan^{-1}\left ( \frac{\alpha (1-\tau^2)\sin\theta}{\tau(1+\alpha^2)-\alpha (1+\tau^2)\cos\theta} \right )
\end{aligned}
\end{equation}

As mentioned earlier, $\theta$ is the round-trip phase delay in the rings. By applying differential signal ($\theta = \theta_{DC} \pm \theta_{mod}$) in the rings, one can achieve electro-optic modulation. Here, $\theta_{DC}$ is the ring optical DC bias point, whereas $\theta_{mod}$ is the result of modulating RF input signal. Now, if we assume rings with negligible loss, i.e. $\alpha \approx 1$, set $\phi_{quad} =\frac{\pi}{2}$, and apply differential modulating signal, trigonometric manipulation of $T(\theta)$ and expanding its series around $\theta_{mod}$ gives us Eq. \ref{eq:Pmo1} to Eq. \ref{eq:Pmo1_last}.


\begin{figure*}[!t]
\begin{equation} \label{eq:Pmo1}
P_{M,o}(\theta_{mod}) = P_I T(\theta)= \frac{P_I}{2} - \gamma_1 P_I \theta_{mod} + \gamma_3 P_I \theta_{mod}^3 - \mathcal{O}(\theta_{mod}^5)
\end{equation}

\begin{align}
\gamma_1 &= \frac{(\tau^2-1)}{\tau^2-2\tau \cos(\theta_{DC})+1}
\\
\\
\gamma_3 &= \frac{(\tau^2-1)(2\cos(\theta_{DC})^2 \tau^2 + \cos(\theta_{DC}) \tau^3 + 2\tau^4 + \cos(\theta_{DC}) \tau -8\tau^2 +2)}{3(\tau^2-2\tau \cos(\theta_{DC})+1)^3}
\label{eq:Pmo1_last}
\end{align}

\hrulefill
\vspace*{4pt}
\end{figure*}



\begin{figure}[!h]
\centering
\includegraphics[width=1\columnwidth]{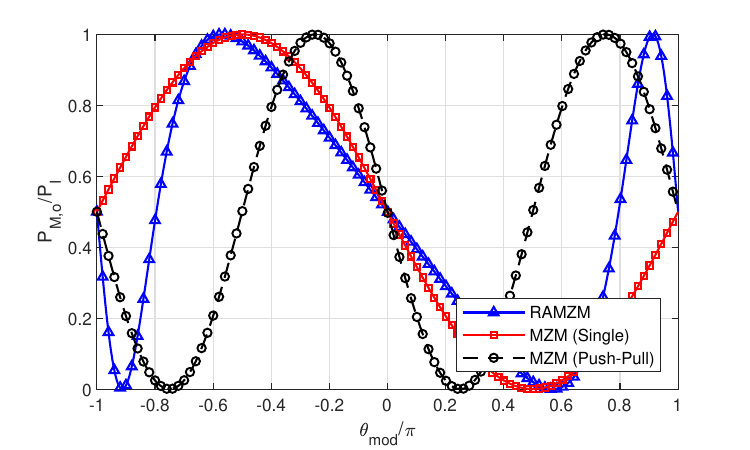}
\caption{RAMZM, MZM (single drive) and MZM (push-pull drive) Optical Transfer Function as a function of voltage induced relative phase shift, $\theta_{mod}$. Here, the RAMZM is biased at $\{\phi_{quad}, \theta_{DC}, \tau \}=\{\frac{\pi}{2},\pi, \frac{1}{2}\}$ and both MZMs are quadrature biased.}
\label{fig:xfr}
\end{figure}

In these equations, $P_I$ is the modulator's optical input power. Upon observing Eq. \ref{eq:Pmo1}, we can see that even-order distortions are entirely eliminated. On the other hand, the third-order nonlinearity ($\gamma_3$) can be eliminated by setting the rings at their anti-resonance point (i.e. optical bias $\theta_{DC} = \pi$) \cite{cardenas2013linearized} and $\tau = \frac{1}{2}$. This is achieved by setting proper coupling coefficients of the rings and tuning the thermal phase shifters of the rings while the driver voltage is set to a common-mode voltage, $V_M$ DC. The differential modulating signals, $v_{m}$, are applied around this common-mode voltage, so that the individual ring drive voltages are $V_M \pm v_{m}$. Therefore, the round-trip phase shift at the upper and lower ring will be $\pi+\theta_{mod}(v_m)$ and $\pi-\theta_{mod}(v_m)$, respectively. This ensures that third-order distortion is suppressed, leaving the modulator with fifth or higher odd-order distortions, which are very minuscule in practical application scenarios. The optical power transfer function of a linearized $(\{\phi_{quad}, \theta_{DC}, \tau\} = \{\pi/2, \pi, \frac{1}{2}\})$ RAMZM along with the transfer functions of MZMs (both single and push-pull drive \cite{zhu2015design}) are plotted in Fig. \ref{fig:xfr}. As can be seen, the RAMZM is significantly linear compared to the MZMs when biased at this regime.

\subsection{RAMZM Link Gain and Slope Efficiency}

Now, lets express the phase modulation as $ \theta_{mod} = \frac{\pi v_{m}}{V_{\pi}}$. Here, $V_{\pi}$ is defined as the voltage required to achieve $\pi$ phase shift in the given length of the phase modulator. From here, we obtain the RAMZM small signal output optical power as \cite{cox2006analog}-

\begin{equation} \label{eq:Pmo2}
P_{M,o} = \frac{\pi \gamma_1 P_I v_m}{V_{\pi} L}
\end{equation}

Here, the term `L' is used to include the effects of loss. When lossy impedance matching \cite{cox2006analog} at the input side is implemented, the voltage applied to the modulator is related to the available power by-

\begin{equation} \label{eq:Psa_vm}
    v_m = \frac{v_s}{2} = \sqrt{P_{s,a}R_s}
\end{equation}

where, $P_{s,a}$ is the input RF power. Consequently, incremental modulation efficiency is derived as \cite{cox2006analog} -

\begin{align} \label{eq:G_m_small_signal}
    G_M &= \frac{P_{M,o}^2}{P_{s,a}} 
        = \frac{s_{rmz}^2}{R_s}= \Big[\frac{\pi \gamma_1 P_I R_s}{V_{\pi} L} \Big ]^2 \frac{1}{R_s} 
\end{align}

where $s_{rmz} \triangleq \frac{P_{M,o}}{v_m} = \frac{\pi \gamma_1 P_I R_s}{V_{\pi} L} = \frac{\pi P_I R_s}{3V_{\pi} L}$ (when $\tau = \frac{1}{2}, \gamma_1 = \frac{1}{3}$) is the RAMZM slope efficiency. The effective modulation efficiency can be increased by increasing the input optical power $P_I$ (from a CW laser), reducing the insertion loss $L$, or decreasing modulator's $V_{\pi}$. To put that in perspective, MZM operating in singe and push-pull drive has the slope efficiencies of $s_{mz} = \frac{\pi P_I R_s}{2V_{\pi} L}$ and $s_{mz} = \frac{\pi P_I R_s}{V_{\pi} L}$, respectively.

Photo-detection in silicon-based PICs is realized using on-chip Germanium detectors \cite{byrd2017mode}. The detector is either interfaced with a passive load $R_L$, or a transimpedance amplifier (TIA). In this work, it is assumed that a matching resistor is present in parallel to the photodetector to match the output load resistance. Therefore, the detected power at the load is $P_{L} = \frac{1}{4}i_d^2 R_L = (\frac{1}{2}r_d \cdot P_{o,d})^2 R_L$ and thus incremental detection efficiency is \cite{cox2006analog}-

\begin{equation} \label{eq:G_det}
	G_{det} \triangleq \frac{P_{L}}{p_{o,d}^2} = \frac{(\frac{1}{2}r_d P_{o,d})^2 R_L }{P_{o,d}^2} = \frac{1}{4}r_d^2 R_L
\end{equation}
where $r_d$ is the detector responsivity. 

To be able to find the gain of the entire link, the product of the modulator and detector incremental efficiencies must be evaluated. If there is an optical loss/gain element between modulator and the detector, it can be modeled as  $\frac{p_{o,d}^2}{P_{M,o}^2}=g^2$. Assuming $R_s=R_L=R$, the entire link gain can be expressed as-

\begin{equation} \label{eq:G_links}
	G = \frac{P_{M,o}^2}{P_{s,a}}\cdot \frac{p_{o,d}^2}{P_{M,o}^2}\cdot\frac{P_{L}}{p_{o,d}^2} = \Big[\frac{g\pi \gamma_1 P_I R r_d}{2V_{\pi} L} \Big ]^2
\end{equation}

Now, let's develop a deeper understanding by conducting some numerical analysis. To evaluate the link gain numerically, the link parameters presented in Table \ref{tab:link_param} are used. These parameters are typical of traditional RF photonic links.

\begin{table}[hbt!]
\caption{Analog Optical Link Parameters}
\label{tab:link_param}
\begin{center}
\begin{tabular}{|l|c|c|}
\hline

Parameter & Value & Unit\\

\hline
\hline
Laser input power ($P_I$) & +19 & dBm\\
\hline
Laser Relative intensity noise (RIN) & -165 & dB/Hz\\
\hline
Modulator $V_{\pi}$ & 8 & V\\
\hline
Modulator Insertion loss (IL) & 6 & dB\\
\hline
Source Impedance ($R_s$) & 50 & Ohm\\
\hline
Detector responsivity ($r_{d}$) & 0.60 & A/W\\
\hline
Load Impedance ($R_L$) & 50 & Ohm\\
\hline
Noise BW & 1 & Hz\\
\hline
Optical Attenuator / Gain Element & None & -\\
\hline
\end{tabular}
\end{center}
\end{table}

\begin{figure}[!h]
\centering
\includegraphics[width=1\columnwidth]{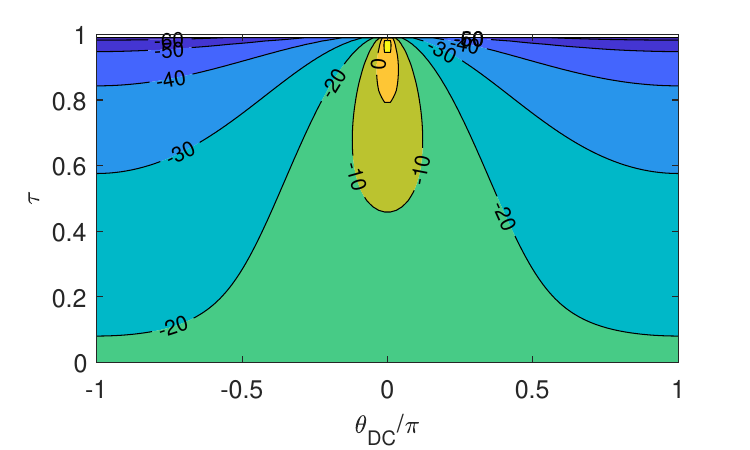}
\caption{Contour plot of RAMZM-based analog optical link gain at different bias conditions $(\theta_{DC}, \tau)$ while keeping $\phi_{quad} = \frac{\pi}{2}$. The link parameters employed are listed in Table \ref{tab:link_param}.}
\label{fig:linkgain3d}
\end{figure}

Fig. \ref{fig:linkgain3d} shows the contour plot for the RAMZM link gain, $G$, for various bias conditions $(\theta_{DC}, \tau)$.

\subsubsection{Improving RAMZM Gain}

\begin{figure}
\centering
\includegraphics[width=1\columnwidth]{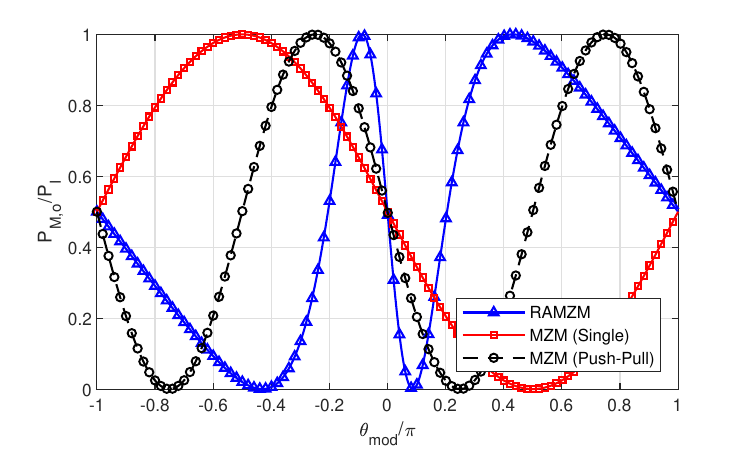}
\caption{RAMZM, MZM (single drive) and MZM (push-pull drive) Optical Transfer Function as a function of RF voltage induced relative phase shift, $\theta_{mod}$. Here, the RAMZM is biased at $\{\phi_{quad}, \theta_{DC}, \tau \}=\{\frac{\pi}{2},0,\frac{1}{2}\}$ and both MZMs are quadrature biased.}
\label{fig:highgain}
\end{figure}

It's evident from the previous analysis and Fig. \ref{fig:xfr} that the linear response of RAMZM comes at the cost of the small-signal power gain. While RAMZM provides excellent distortion cancellation upto 5th order, its slope efficiency is only 67\% and 33\% of that of an MZM driven in single and push-pull manner, respectively. To improve the slope efficiency of the RAMZM, the derivative of the transfer function with respect to $\theta$ can be evaluated to find the bias point where slope efficiency is the maximum for a given $\tau$. Here, we propose a biasing scheme where $\theta_{DC} = 0$, instead of $\pi$ while keeping $\phi_{quad} = \frac{\pi}{2}$. If we choose $\tau = \frac{1}{2}$ (i.e $\gamma_1 = 3$), the incremental modulation efficiency can be calculated as-

\begin{align} \label{eq:G_m_small_signal_high_slope}
    G_M &= \Big[\frac{3 \pi P_I R_s}{V_{\pi} L} \Big ]^2 \frac{1}{R_s} 
\end{align}

where $s_{rmz} \triangleq \frac{3\pi P_I R_s}{V_{\pi} L}$. The optical power transfer function of this RAMZM $(\{\phi_{quad}, \theta_{DC}, \tau\} = \{\pi/2, 0, \frac{1}{2}\})$ along with the transfer functions of MZMs (both single and push-pull drive) are plotted in Fig. \ref{fig:highgain}. As can be seen, the RAMZM has significantly higher slope efficiency compared to the MZMs. Here, slope efficiency is 3$\times$ better than that of MZIs driven in push-pull manner and 6$\times$ better than that of MZMs operated in single drive \cite{cox2006analog,marpaung2009high}. This means, for the same link gain, this biasing scheme will allow upto 6$\times$ reduction in phase shifter length compared to MZMs, enabling lumped drive, smaller form factor, smaller energy footprint and excellent noise performance. However, this significantly higher gain and noise performance comes at the cost of linearity since $\gamma_3$ is no longer 0 in this biasing condition.

\subsection{RAMZM Link Noise Figure}

The noise figure (NF) for the entire link is given by \cite{cox2006analog}-

\begin{equation} \label{eq:NF2}
	NF =10 \cdot log_{10} \Big(\frac{N_{out}}{G N_i} \Big) 
\end{equation}

where $N_{out}$ is the total noise power at the link output, $N_i$ is the input noise, and $G$ is the link gain. The input noise power is due to the source resistance and given by $N_i=kT \Delta f$ \cite{cox2006analog}. The noise components of the link can be summarized as \cite{cox2006analog}- 

\begin{enumerate}
\item Laser relative intensity noise (RIN), $\overline{i_{rin}^2} \Delta f$ 
\item Detector shot noise, $\overline{i_{sn}^2} \Delta f$ 
\item Thermal noise from the load matched resistance of the detector ($R_L=R_{pd}$), $kT \Delta f$ 
\end{enumerate}

RIN of the laser source is expressed in dB/Hz and the resulting RIN noise current spectral density (CSD) is given by $\overline{i_{RIN}^2} = \overline{I_D}^2 10^{\frac{RIN}{10}} $ where $\overline{I_D} = r_d P_{av} = \frac{r_d P_I}{2 L}[1+cos(\phi_{quad})]$ is average detector current \cite{cox2006analog}. It is important to note that $P_{av} = \frac{P_I}{2 L}[1+cos(\phi_{quad})]$ comes from Eq. \ref{eq:Pmo1}, weighted by optical loss L. The detector shot noise current spectral density is described as $\overline{i_{sn}^2} = 2q \overline{I_D}$. Thus, RIN and shot noise CSD depend upon $\overline{I_D^2}$ and $\overline{I_D}$ respectively. 

When lossy impedance matching is implemented at the input and output side, the total noise for the link is derived as


\begin{equation} \label{eq:NF_link1}
	N_{out}= \frac{1}{4}(\overline{i_{rin}^2} + \overline{i_{sn}^2})\Delta f R_L + (1+G)kT \Delta f 		
\end{equation}

Consequently, the NF of the link is expressed as

\begin{equation} \label{eq:NF_primary}
	NF = 10\cdot log_{10} \Big[1 + \frac{R_L}{4kTG} \Big(I_D^2 10^{\frac{RIN}{10}} + 2q \overline{I_D} \Big) + \frac{1}{G} \Big]	
\end{equation}

\begin{figure}[!h]
\centering
\includegraphics[width=1\columnwidth]{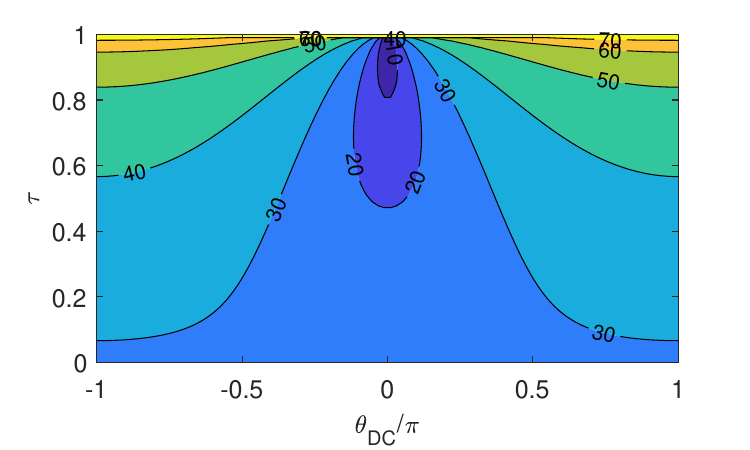}
\caption{Contour plot of RAMZM-based analog optical link NF at different bias conditions $(\theta_{DC}, \tau)$ while keeping $\phi_{quad} = \frac{\pi}{2}$. The link parameters employed are the same as in Table \ref{tab:link_param}.}
\label{fig:nfplot3d}
\end{figure}

The corresponding plot of NF for the different bias conditions $(\theta_{DC}, \tau)$ is illustrated in Fig. \ref{fig:nfplot3d}. Here, it is evident that for $\theta_{DC} = 0$, NF is minimized due to the large link gain.

The corresponding plot of NF for the different bias conditions $(\theta_{DC}, \tau)$ is illustrated in Fig. \ref{fig:nfplot3d}. Here, it is evident that for $\theta_{DC} = 0$, NF is minimized due to the large link gain and lower optical output power. One can observe that the link NF is rather high due to the detector noise, as $\overline{I_D}$ depends upon $\frac{P_I}{2 L}[1+cos(\phi_{quad})]$. This can be improved by biasing the arms of the modulator at different angle, $\phi_{quad}> \frac{\pi}{2}$ \cite{cox2006analog,marpaung2009high}, or by filtering out the optical carrier \cite{lagasse1994optical,marpaung2009high} using an on-chip SiP filter \cite{shawon2020rapid}.


\section{RAMZM Linearity} \label{sec:link_distortion_analysis}

\begin{figure}[!h]
\centering
\includegraphics[width=1\columnwidth]{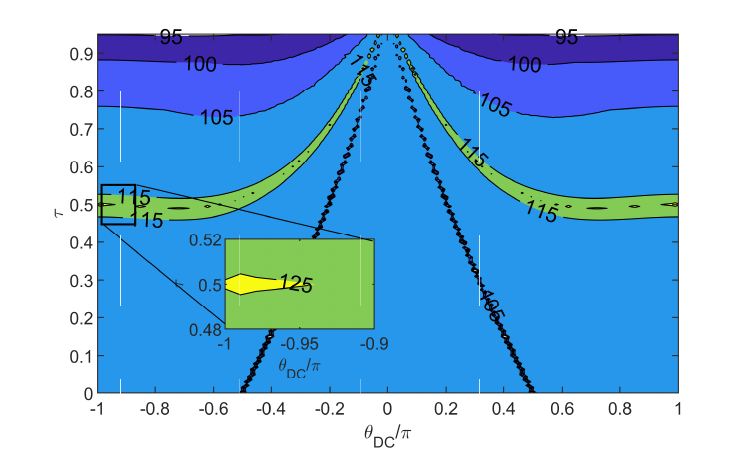}
\caption{Contour plot of RAMZM-based analog optical link SFDR in $dB.Hz^{\frac{2}{3}}$ at different bias conditions, keeping $\phi_{quad} = \frac{\pi}{2}$. The link parameters employed are listed in Table \ref{tab:link_param}.}
\label{fig:sfdr3dfinal}
\end{figure}

To provide a comprehensive view of the link gain and total noise, SFDR is evaluated and plotted in Fig. \ref{fig:sfdr3dfinal} using the method described in \cite{kundert2002accurate} and link parameters listed in Table \ref{tab:link_param}. Here, we see that the highest SFDR of $\sim 127$ dB.Hz$^{2/3}$ is obtained when the rings are biased in anti-resonance and $\tau=\frac{1}{2}$. As ring bias are tuned away from anti-resonance, a narrow window for coupling ratio $\tau$ exists for highest SFDR. The generated contours in Figs. \ref{fig:linkgain3d}, \ref{fig:nfplot3d} and \ref{fig:sfdr3dfinal} can be used together to make design trade-offs among the link gain, NF and linearity.

\subsection{Linearity of RAMZM and MZM}

\begin{figure}[!h]
\centering
\includegraphics[width=1\columnwidth]{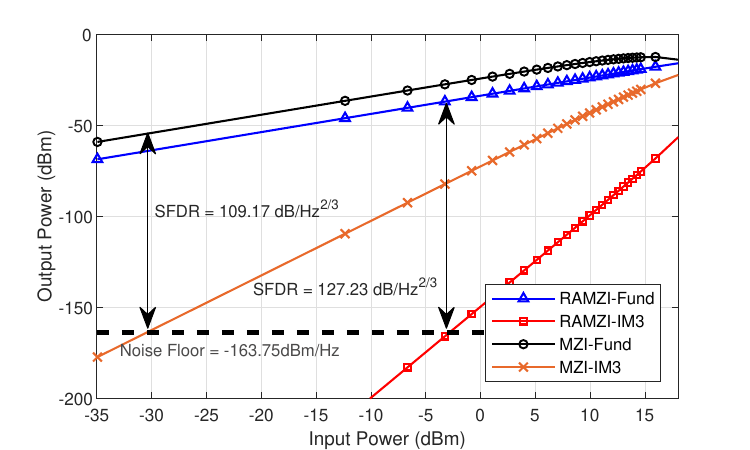}
\caption{SFDR performance of RAMZM and MZM in a Photonic Link. Here, RAMZM and MZM bias conditions are $\{\phi_{quad}, \theta_{DC}, \tau \}=\{\frac{\pi}{2},\pi, \frac{1}{2}\}$ and $\{\phi_{quad}\}=\{\frac{\pi}{2}\}$, respectively. Also, the link parameters are from Table \ref{tab:link_param}.}
\label{fig:ramzmvsmzmsfdr}
\end{figure}

As discussed before, MZM suffers from higher nonlinearity due to its sinusoidal transfer function. To quantify the linearity of both MZM and linearized RAMZM, $SFDR_3$ is calculated in Fig. \ref{fig:ramzmvsmzmsfdr}. Here, RAMZM and MZM have SFDR values of 127.23 and 109.17 $dB.Hz^{\frac{2}{3}}$ at 1GHz, respectively. This is 18 dB improvement over regular MZMs, which makes RAMZM very attractive for high-performance RF photonic applications.

\subsection{Linearity of Gain Enhanced (GE) RAMZM}

\begin{figure}[!h]
\centering
\includegraphics[width=1\columnwidth]{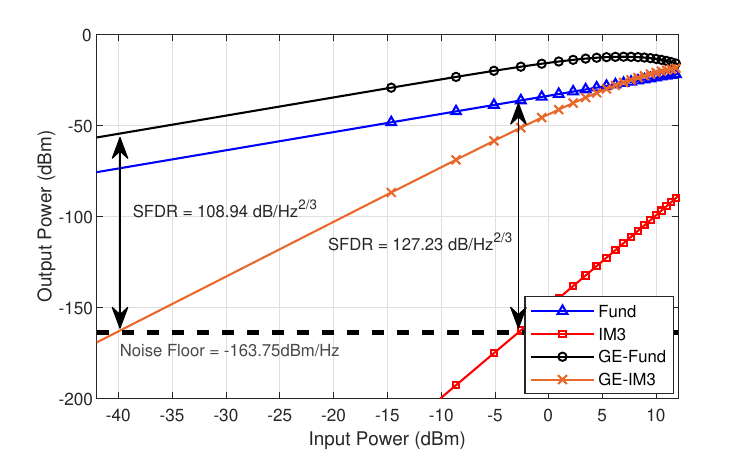}
\caption{Comparison of SFDR performance of RAMZMs in a Photonic Link with different bias conditions. Here, Linearized and Gain Enhanced (GE) RAMZM bias conditions are $\{\phi_{quad}, \theta_{DC}, \tau \}=\{\frac{\pi}{2},\pi, \frac{1}{2}\}$ and $\{\frac{\pi}{2},0, \frac{1}{2}\}$, respectively. Also, the link parameters are from Table \ref{tab:link_param}.}
\label{fig:ramzivsramzi}
\end{figure}

In previous section, we proposed a new biasing scheme where the slope efficiency of RAMZM can be made very high. However, as previously discussed, this gain enhancement comes at the expense of linearity since $\theta_{DC}$ has to be shifted from $\pi$ towards 0. In Fig. \ref{fig:ramzivsramzi}, we show that the SFDR of GE RAMZM is about 108.94 $dB.Hz^{\frac{2}{3}}$ at 1GHz. It is interesting to note that the GE RAMZM provides 6$\times$ improvement in slope efficiency (when $\tau = \frac{1}{2}$) over MZM while still providing a similar SFDR performance as an MZM.

\section{Design and Fabrication of Ring Assisted Mach Zehnder Modulator} \label{sec:secDFP}

\subsection{Modulator Topology}

\begin{figure}[htbp]
\centering
\includegraphics[width=1\linewidth]{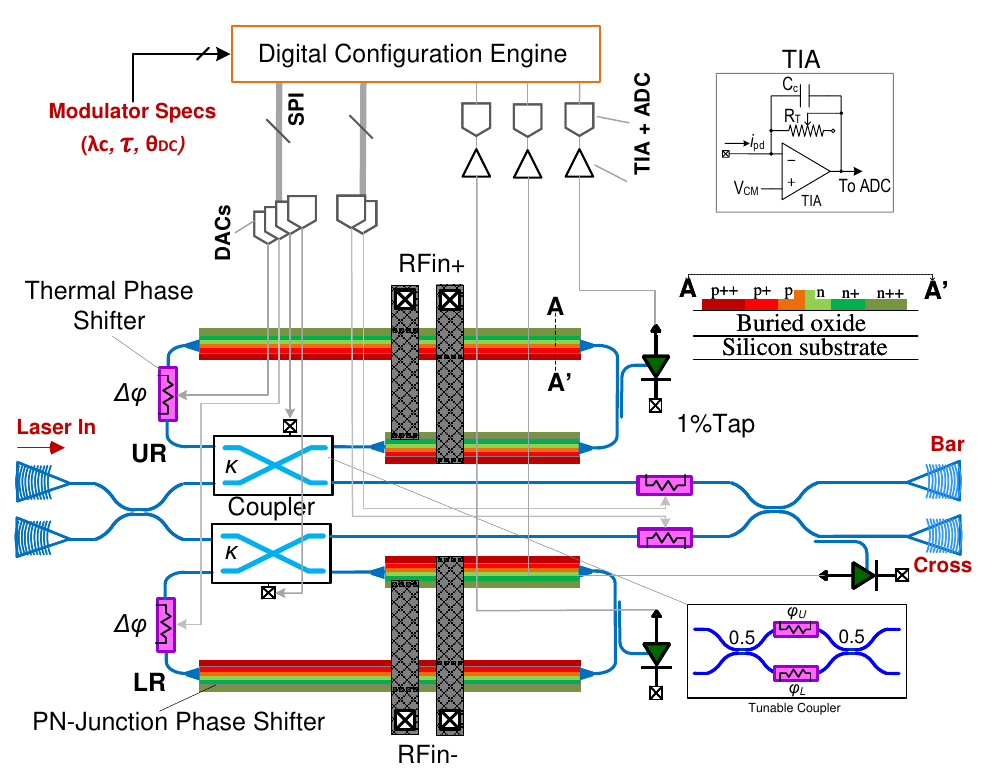}
\caption{Schematic diagram of the RAMZM and its interfacing with the electronic backend.}
\label{fig:RamzmFullSch}
\end{figure}

Fig. \ref{fig:RamzmFullSch} portrays the schematics of the actual RAMZM Photonic Integrated Circuit (PIC) realized in this work. A continuous-wave (CW) laser is the input, which gets channeled into both the arms of a Mach Zehnder Interferometer (MZI). Each arm of the MZI is loaded with a ring (UR/LR) through a tunable coupler (UR-C/LR-C) and quadrature phase-shifter (Q-PS). Both rings include a high-speed phase modulator and a microheater for resonance adjustment. The phase modulators need differential inputs to operate in a linear regime. The tunable couplers are realized using $2\times2$ MZI switch (shown in the inset of \textbf{Fig. \ref{fig:RamzmFullSch}}).

To keep track of the ring resonance, a 1\% tap followed by an on-chip Ge-photodetector (PD) is employed for each ring. Despite increasing the passband loss of the modulator, these taps are vital for the automatic software reconfiguration algorithm \cite{choo2018automatic}. Moreover, a 10\% monitor tap is utilized at the cross port of the modulator for quadrature bias monitoring. The calibration and tuning algorithm operates automatically with the assistance of on-chip thermo-optic phase-shifters or microheaters and monitor taps.

\subsection{High-Speed Phase Modulator Design}
\begin{figure}[!tbh]
\centering
\includegraphics[width=1\columnwidth]{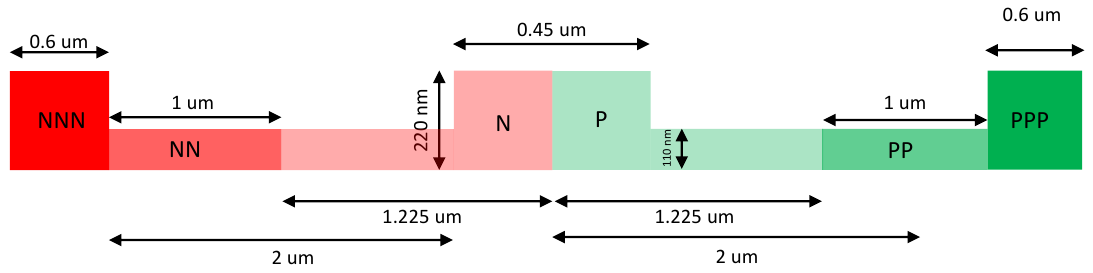}
\caption{Cross-section of the high-speed phase modulator.}
\label{fig:PhaseShifterCrossSection}
\end{figure}

As mentioned before, each ring of the RAMZM is loaded with a high-speed phase modulator. These modulators are basically lateral PN junctions operating in reverse bias and utilize the plasma dispersion effect to induce voltage-dependent phase shift to the propagating light. For this modulator, we designed a 220nm $\times$ 450nm rib waveguide with $\sim$110nm etched silicon. As shown in Fig. \ref{fig:PhaseShifterCrossSection}, the silicon rib waveguide is lightly doped to form a lateral p-n junction at the center. Moderate doping was used in the region between the waveguide and the highly doped ohmic contact region. This region (moderate doping) was 1.225$\mu$m away from the center of the waveguide and provided a balanced tradeoff between the series resistance of the modulator and excess free carrier absorption. Although longer phase modulators have smaller $V_{\pi}$ which in turn provides better link gain, for this work only 1555um long modulator is considered to allow lumped drive and higher RC limited bandwidth.

\begin{figure}[!tbh]
\centering
\includegraphics[width=1\columnwidth]{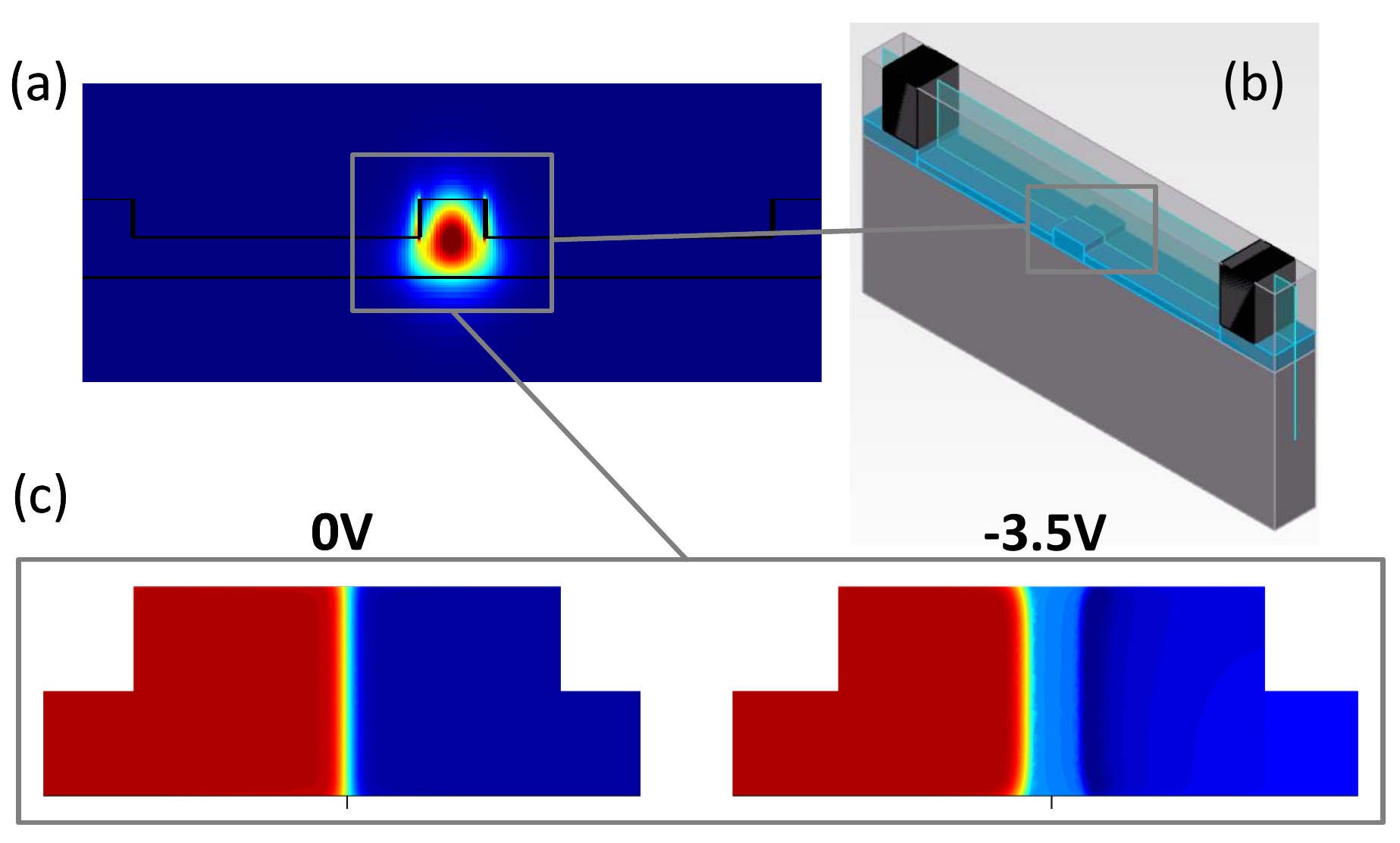}
\caption{(a) Optical mode profile of the rib waveguide. (b) Rib waveguide structure created in Ansys Lumerical CHARGE platform to extract the charge profile in the laternal PN junction. (c) Charge (n) distribution at 0v and -3.5V bias.}
\label{fig:ramzmPhaseModSim}
\end{figure}

The phase modulator's simulation and verification were performed using Ansys Lumerical's MODE and CHARGE solvers, as depicted in Fig. \ref{fig:ramzmPhaseModSim}. The doping concentrations utilized in these calculations were back-calculated from the sheet resistance supplied in the PDK. From the simulation, the effective index ($n_{eff}$) and the DC capacitance relative to the reverse bias voltage were extracted. Furthermore, the series resistance (from N-side contact to P-side contact) was determined to be 4.13 $\Omega$. This resistance value, in conjunction with the extracted capacitance, allowed the computation of the RC-limited bandwidth of the phase modulator.

\begin{figure}[!tbh]
\centering
\includegraphics[width=1\columnwidth]{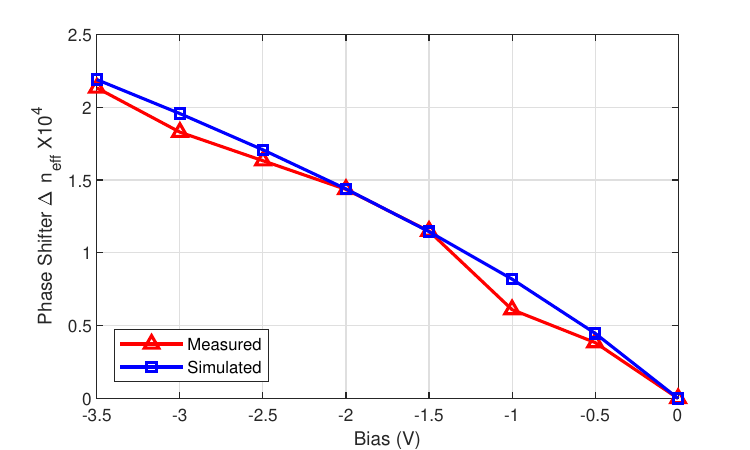}
\caption{Simulated and measured $n_{eff}$ of the PN modulator as a function of bias voltage.}
\label{fig:phaseModulatorNeff}
\end{figure}

\begin{figure}[ht!]
\centering 
(a)
\includegraphics[width=1\linewidth]{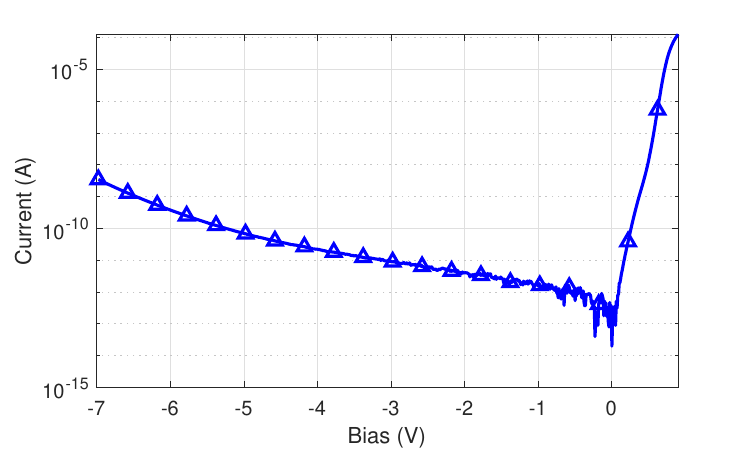}
(b)
\includegraphics[width=1\linewidth]{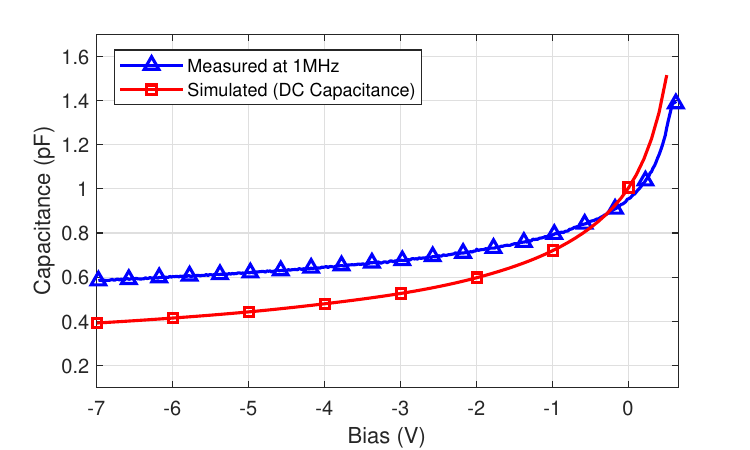}
(c)
\includegraphics[width=1\linewidth]{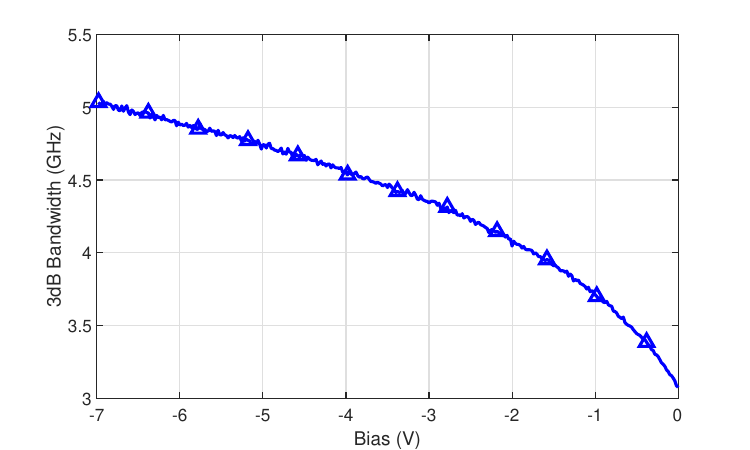}
\caption{(a) Measured IV curve of the Phase modulator. (b) Simulated and measured capacitance of the modulator. (c) RC limited bandwidth (50 $\Omega$ system) of the modulator derived from the measured capacitance.}
\label{fig:phaseModulatorEData}
\end{figure}

Experimental extraction was also conducted for the $n_{eff}$ and capacitance of the modulator in relation to voltage. The phase modulator was subjected to various reverse bias voltages and the corresponding phase shift was then recorded. This data from the phase shifter enabled the calculation of the change in the effective index relative to zero bias. The simulated and experimentally extracted $n_{eff}$ data is presented in Fig. \ref{fig:phaseModulatorNeff}. Conversely, for capacitance extraction, a Multi-Frequency Capacitance Measurement (MFCMU) Unit was utilized and capacitance was measured at 1MHz. The same equipment was used to measure the I-V characteristic of the phase modulator. By using the extracted capacitance and resistance values, the RC-limited bandwidth was also calculated, as illustrated in Fig. \ref{fig:phaseModulatorEData}.

\subsection{RAMZM PIC Layout and Fabrication} 

The RAMZM PIC has been fabricated in AIM Photonics foundry’s 300mm wafer Active PIC process. Process-optimized grating couplers (GC) are used to get light in and out of the PIC. 220nm x 480nm silicon waveguides were utilized for interconnecting the optical components. Tunable couplers were implemented using a 2x2 thermo-optic switch, which is essentially a balanced MZ structure with microheaters on each arm. The microheater thermo-optic time constants are on the order of $\sim$15$\mu$s. The 1$\%$ and 10$\%$ monitor taps used in this design are on-chip Ge photodetectors (PD) with couplers and optical terminations.

\begin{figure}[!tbh]
\centering
\includegraphics[width=0.9\columnwidth]{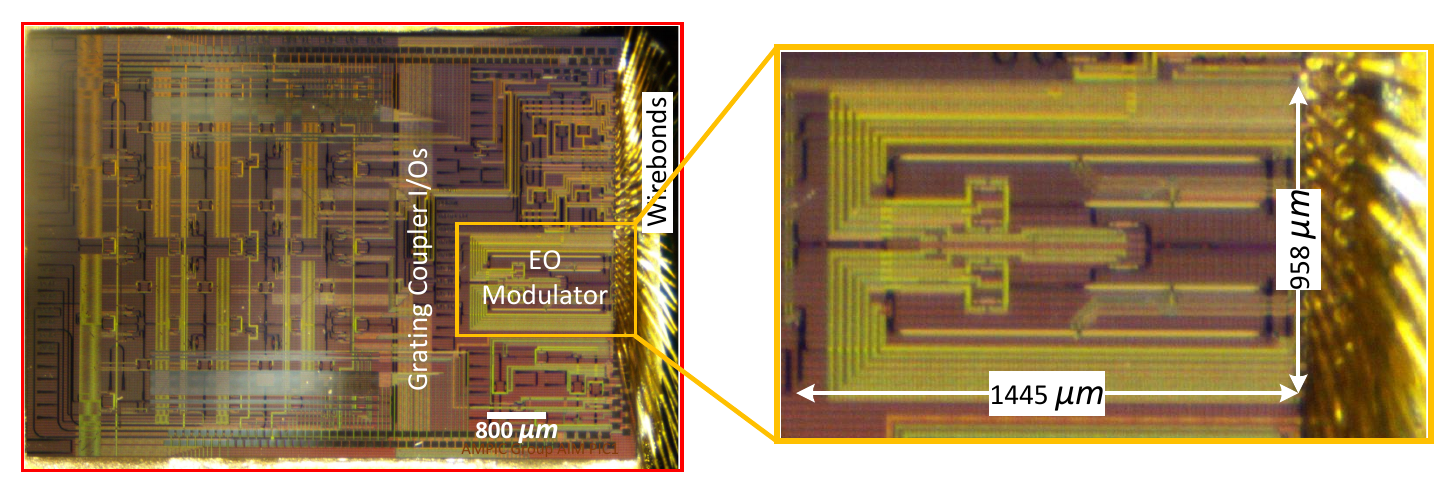}
\caption{Chip micrograph of the RAMZM fabricated in AIM Photonics SiP process.}
\label{fig:ramzmdie}
\end{figure}

The rings were routed with $>$15$\mu$m bend radius for lower bending loss. The total physical length of the rings resulted in a free-spectral range (FSR) of around 28 GHz. The monitor taps were realized using 1\% and 10\% couplers, Ge waveguide detectors and waveguide terminations.  The modulator die micrograph is shown in Fig. \ref{fig:ramzmdie}. The modulator core area occupies 1.38mm$^2$ on the chip.

\subsection{Packaging and Electronic Back-end} 
\begin{figure}[!tbh]
\centering
\includegraphics[width=0.8\columnwidth]{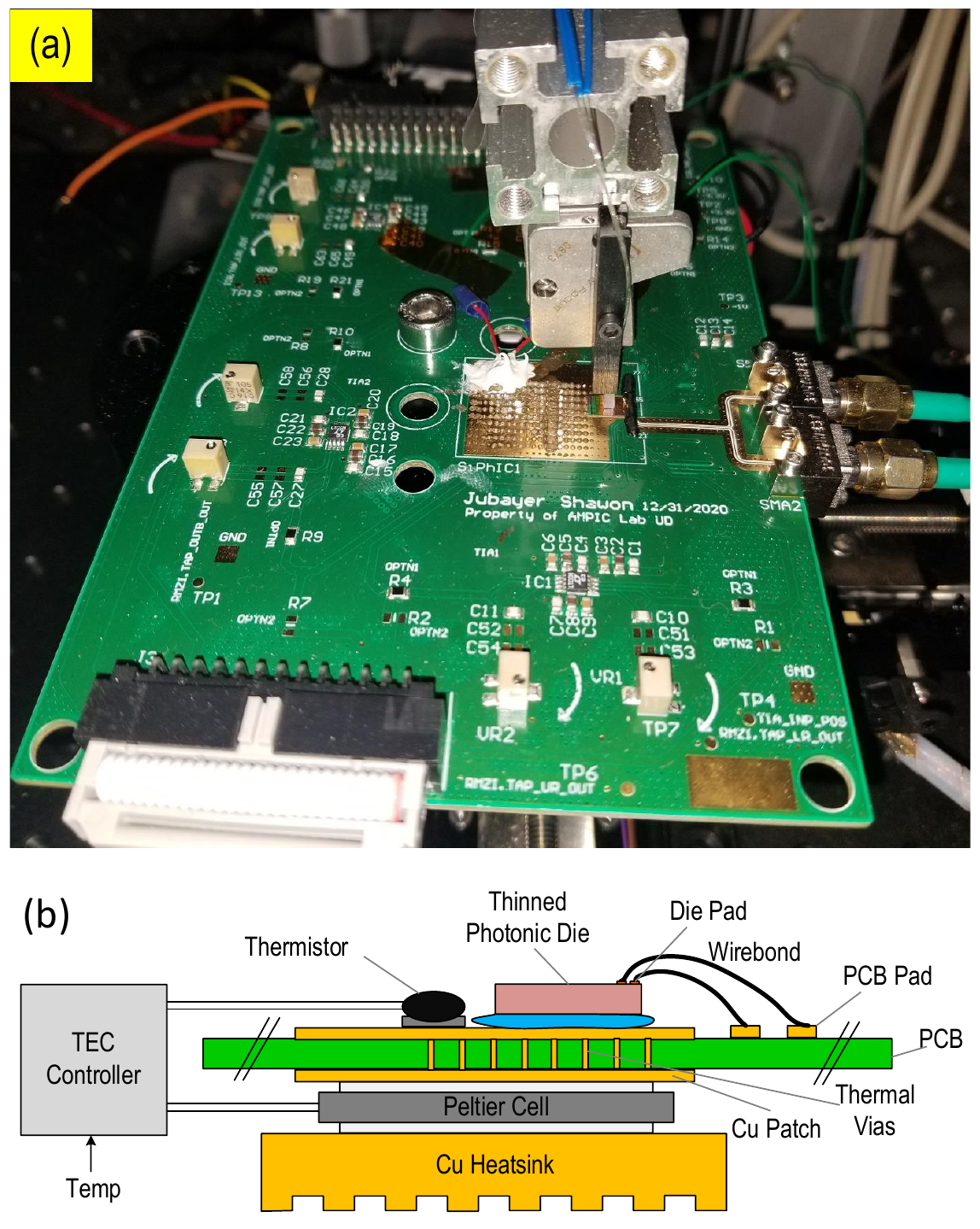}
\caption{(a) The chip is attached and wire-bonded onto a custom designed printed circuit board (PCB) using Chip-On-Board (COB) packaging.  The PCB contains transimpedance amplifiers (TIAs) and ribbon cable interface to another electronic board with DACs and ADCs.  (b) The COB assembly is co-packaged with a thermistor on the top-side and a Peltier cell on the back side of the PCB with an external TEC controller board in a closed loop.}
\label{fig:ramzmpcb}
\end{figure}

The fabricated PIC die was polished down to 150$\mu m$ and packaged in a chip-on-board (COB) assembly as shown in Fig. \ref{fig:ramzmpcb}(a \& b). A Peltier cell and thermistor were used along with a thermo-electric cooler (TEC) controller in a closed-loop to stabilize the temperature of the chip and to minimize the thermal crosstalk among the on-chip tuning elements. The combination of die thinning and TEC at the bottom of the die provides effective thermal isolation by creating a prominent thermal gradient in the vertical direction \cite{choo2018automatic}. 

The modulator RF feed comprises of impedance controlled single-ended/differential Coplanar Waveguides (CPWs) and end-launch connectors. The electrical pads were placed on two rows on the East edge of the PIC which were wire-bonded with two rows of PCB pads. The optical monitors were connected to the on-board transimpedance amplifiers (TIAs), whose outputs were interfaced with commercial off the shelf (COTS) 16-bit analog-to-digital converters (ADCs) using a ribbon cable. These pads provide 16-bit digital-to-analog converters (DACs) interfaces to the microheaters. All the DACs and ADCs communicate via SPI interface with a microcontroller, which provides a software abstraction to the algorithm code.

\section{In-Situ Component Parameter Extraction and Reconfiguration Algorithm}

To operate the RAMZM modulator at its most linear region, precise bias conditions must be enforced: $\{\phi_{quad}, \theta_{DC}, \tau \}=\{\frac{\pi}{2},\pi, \frac{1}{2}\}$. However, variations in process, voltage, and temperature (PVT) make it challenging for the fabricated chip to consistently maintain the intended operating point. Additionally, users may desire to operate the modulator at different bias points or center wavelengths ($\lambda_c$). Therefore, in-situ component parameter extraction and reconfiguration capability are essential.

For the RAMZM modulator, achieving full reconfigurability and tunability involves adjusting coupler coupling coefficients and aligning ring anti-resonances to user-specified center wavelength ($\lambda_c$). We have developed a systematic algorithm to extract the desired coupling coefficients of tunable couplers, which have been presented in \cite{shawonFilter}. This algorithm enables dialing desired coupling coefficient, ensuring $\tau=\frac{1}{2}$ or any other desired value.

In order to align the anti-resonances of the rings in the RAMZM modulator to the user-specified wavelength, a specific procedure is followed. Firstly, the drop port response of the rings is maximized at $\lambda_c$. Subsequently, a phase bias of $\pi$ is applied to detune the rings. This phase detuning ensures that all the anti-resonances of the rings are precisely aligned with the desired wavelength. The comprehensive methodology for dialing the center wavelength and detuning the rings by the desired phase is elaborated upon in detail in \cite{shawonFilter}. This approach guarantees that the desired bias condition of $\theta_{DC} = \pi$ is achieved consistently.

To ensure quadrature bias in the RAMZM modulator, first, both tunable couplers are adjusted to a zero coupling coefficient setting. Subsequently, the quadrature microheater voltage ($v_{quad}$) is swept while simultaneously recording the response of the 10\% monitor tap at the modulator cross port, denoted as $f(v_{quad})$. By analyzing the recorded data, the microheater voltage ($v_{quad}^*$) at which the cross port monitor tap reaches 50\% of its maximum value provides the desired quadrature bias, $\phi_{quad} = \frac{\pi}{2}$. To simplify this procedure, Algorithm \ref{alg:ramzmQaudBias} outlines the steps involved in achieving the quadrature bias in the RAMZM modulator.

\begin{algorithm}
\footnotesize
\caption{Quadrature Biasing}\label{alg:ramzmQaudBias}
\begin{algorithmic}[1]
\State \textbf{Start}
\State \textbf{Sweep} $v_{quad}$ and record cross port response, $f(v_{quad})$
\State \textbf{Set} $v_{quad}^*\gets$ \textbf{Find} voltage $v_{quad}$ when $f=f_{-3dB}(v_{quad})$ 
\State \textbf{End}
\end{algorithmic}
\end{algorithm}

These steps are done iteratively until the system reaches its thermal steady state. Algorithm \ref{alg:ramzmMain_algorithm} summarizes the entire procedure.

\begin{algorithm}[htbp]
\footnotesize
\caption{Modulator Reconfiguration Algorithm}\label{alg:ramzmMain_algorithm}
\begin{algorithmic}[1]
\State \textbf{Start}
\State \textbf{Get} Modulator specifications ($\tau$, $\lambda_c$) from the user
\State \textbf{Set} $n=1$
\State \textbf{while} $n \leq 2$ \textbf{do}
\State \qquad\textbf{Invoke} the \textbf{coupling coefficient ($\tau$)} dialing routine \cite{shawonFilter}
\State \qquad\textbf{Set} the tunable coupler to zero coupling ($\tau_n = 1$ i.e. $\kappa_n = 0$).
\State \qquad\textbf{Set} $n=n+1$
\State \textbf{end while}

\State \textbf{Invoke} the modulator \textbf{quadrature biasing} routine.

\State \textbf{Set} $n=1$
\State \textbf{while} $n \leq 2$ \textbf{do}
\State \qquad\textbf{Run} the \textbf{coupling coefficient ($\tau$)} dialing routine \cite{shawonFilter}
\State \qquad\textbf{Set} the tunable coupler to desired coupling $\tau_n$.
\State \qquad\textbf{Set} $n=n+1$
\State \textbf{end while}

\State \textbf{Set} $n=1$
\State \textbf{while} $\text{err}_j \geq \text{tol}_j$ \textbf{do}
\State \qquad\textbf{while} $\text{err}_i\geq \text{tol}_i$ \textbf{do}
\State \qquad\qquad\textbf{while} $n\leq 2$ \textbf{do}
\State \qquad\qquad\qquad\textbf{Invoke} the \textbf{ring biasing} routine \cite{shawonFilter}
\State \qquad\qquad\qquad\textbf{Set} $n^{th}$ ring at anti-resonance.
\State \qquad\qquad\qquad\textbf{Set} $n=n+1$
\State \qquad\qquad\textbf{end while}
\State \qquad\qquad\textbf{Calculate} $\text{err}_i$
\State \qquad\textbf{end while}
\State \qquad\textbf{Invoke} the modulator \textbf{quadrature biasing} routine.
\State \qquad\textbf{Calculate} $\text{err}_j$
\State \textbf{end while}
\State \textbf{End}
\end{algorithmic}
\end{algorithm}

\section{Experimental Results}

\subsection{Experimental Setup of RAMZM Testing}

\begin{figure}[!tbh]
\centering
\includegraphics[width=\columnwidth]{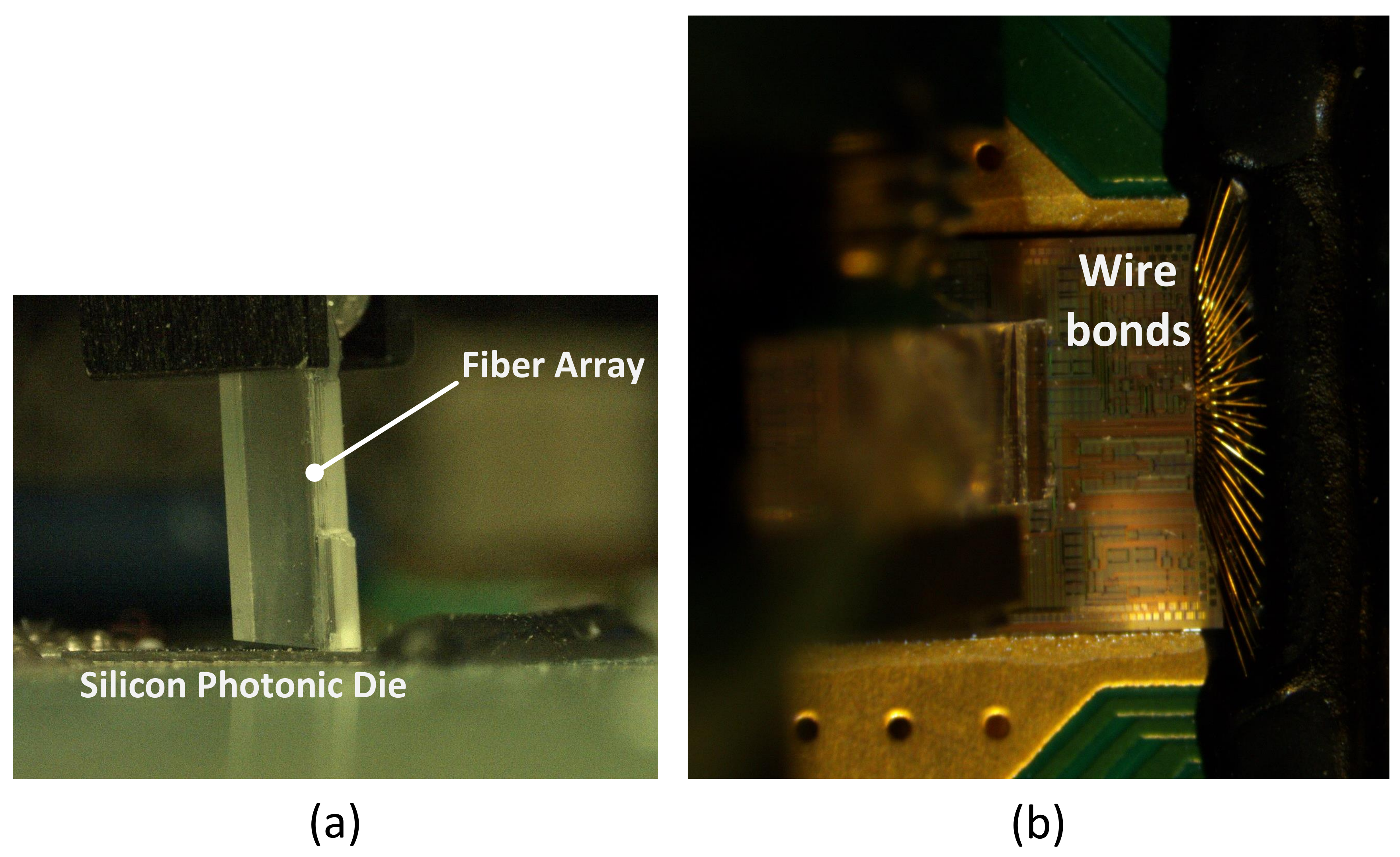}
\caption{(a) A 4-channel polarization maintaining fiber array aligned to the on-chip grating coupler. (b) RAMZM PIC chip-on-board (COB) assembly.}
\label{fig:ramzmAlignment}
\end{figure}

\begin{figure}[!tbh]
\centering
\includegraphics[width=\columnwidth]{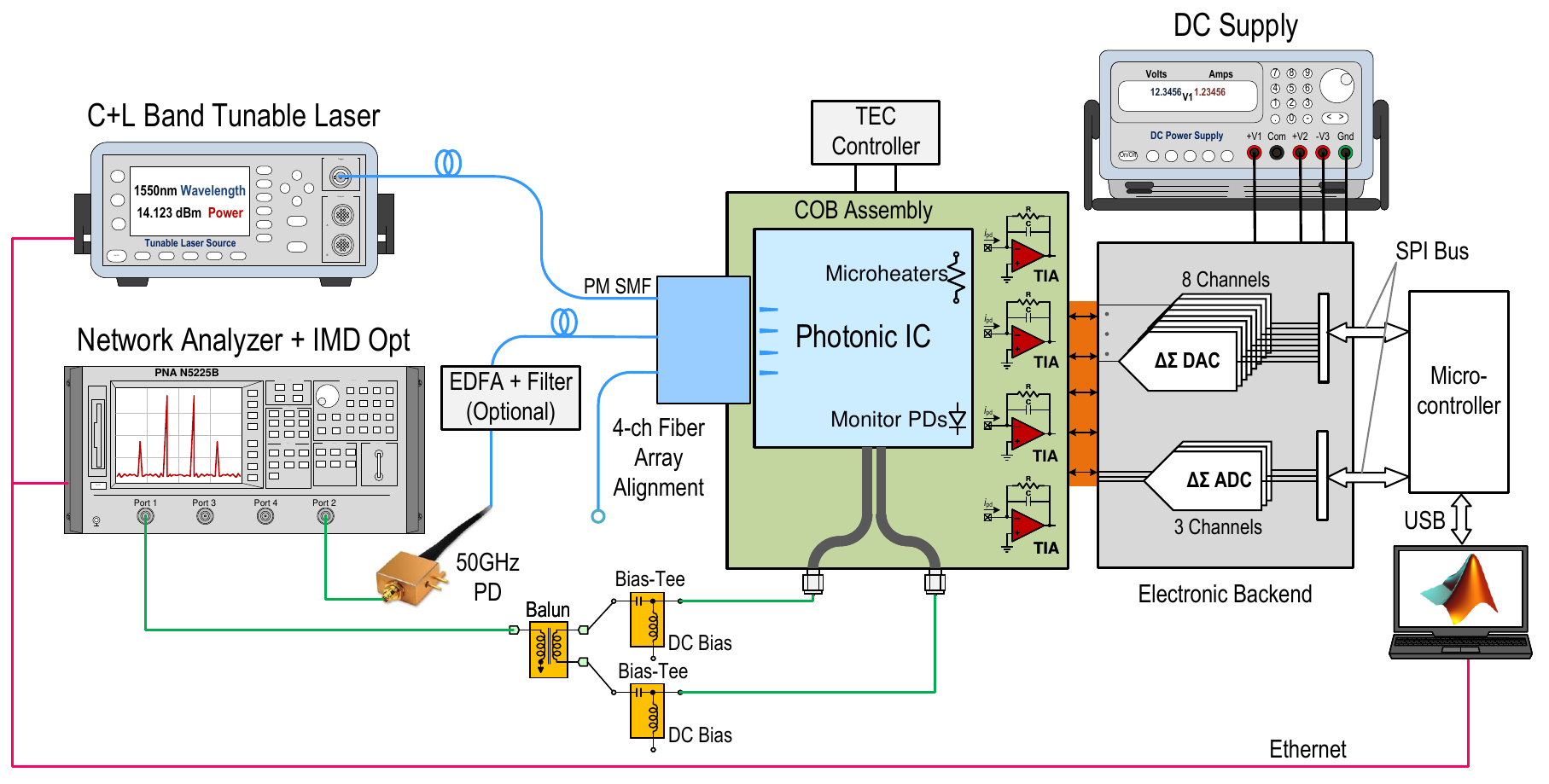}
\caption{Schematics of the experimental setup with the RAMZM PIC aligned to a fiber array, interfaced with the electronic backend and measurement accessories (EDFA, PNA, PD).}
\label{fig:TestSetupSchematics}
\end{figure}

\begin{figure}[!tbh]
\centering
\includegraphics[width=1\columnwidth]{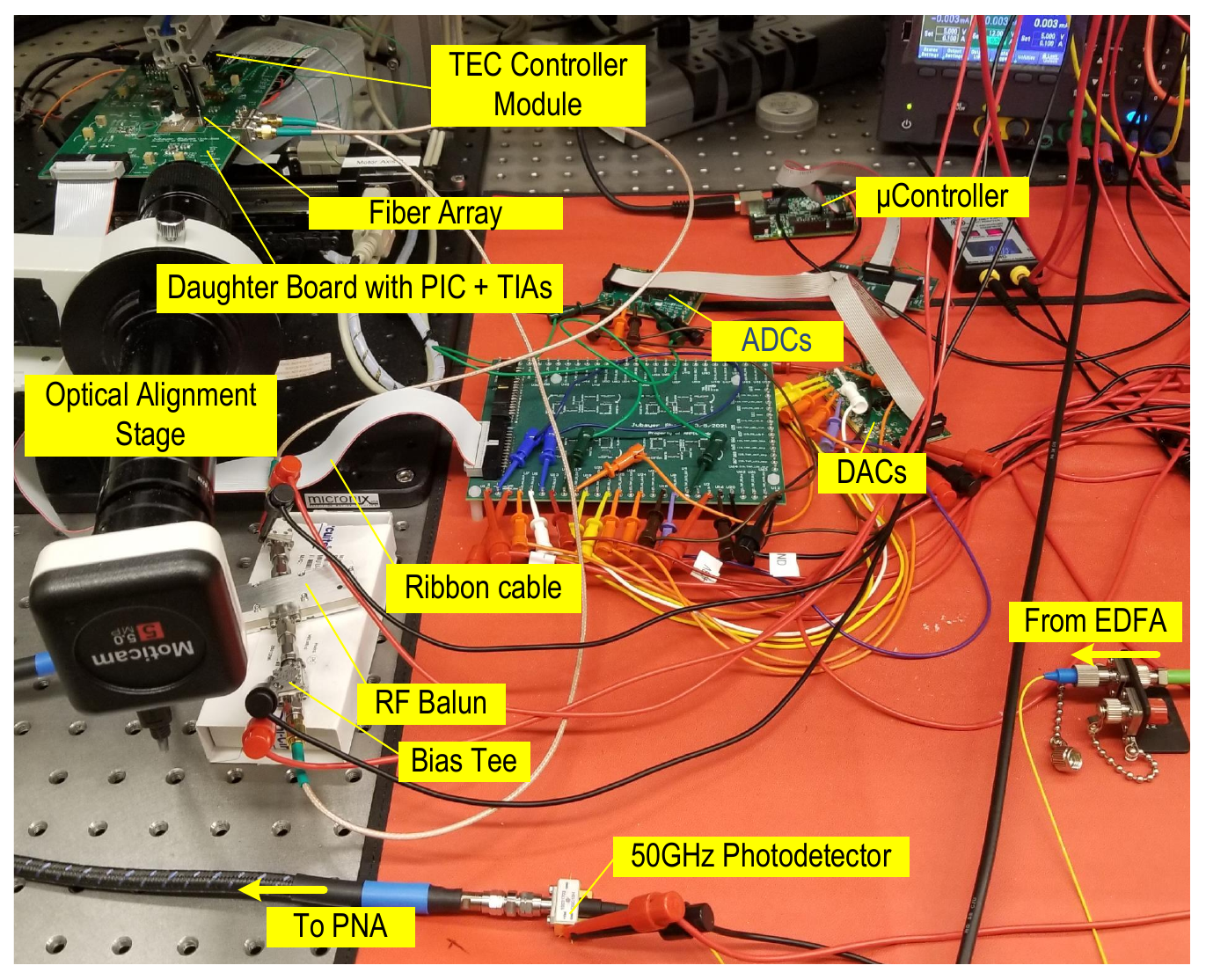}
\caption{Experimental setup with the
RAMZM PIC aligned to a fiber array, interfaced with the electronic backend and measurement accessories (EDFA, PNA, PD).}
\label{fig:RealExperimentalSetup}
\end{figure}

The RAMZM optical I/O is facilitated by 4-ch polarization maintaining fiber array (shown in Fig. \ref{fig:ramzmAlignment}). The fiber array is aligned to the on-chip grating couplers using a semiautomatic alignment stage. The experimental setup consists of the electronic backend for tuning, a Keysight N5225B PNA, EDFA, and a 50GHz photodetector with high linearity (Fig. \ref{fig:TestSetupSchematics}). A single-ended RF input is applied to an RF balun which provides differential output signal pair. The differential signal pass through a bias-tee which provides DC offset to the modulation signal. The differential signals are then applied to one of each ring and the DUT output is recorded using a linear PD (after passing through an EDFA). Fig. \ref{fig:RealExperimentalSetup} shows the actual test setup in the lab.

\subsection{RAMZM Reconfiguration and Bias Tuning}
\begin{figure}[ht!]
\centering
\includegraphics[width=1\linewidth]{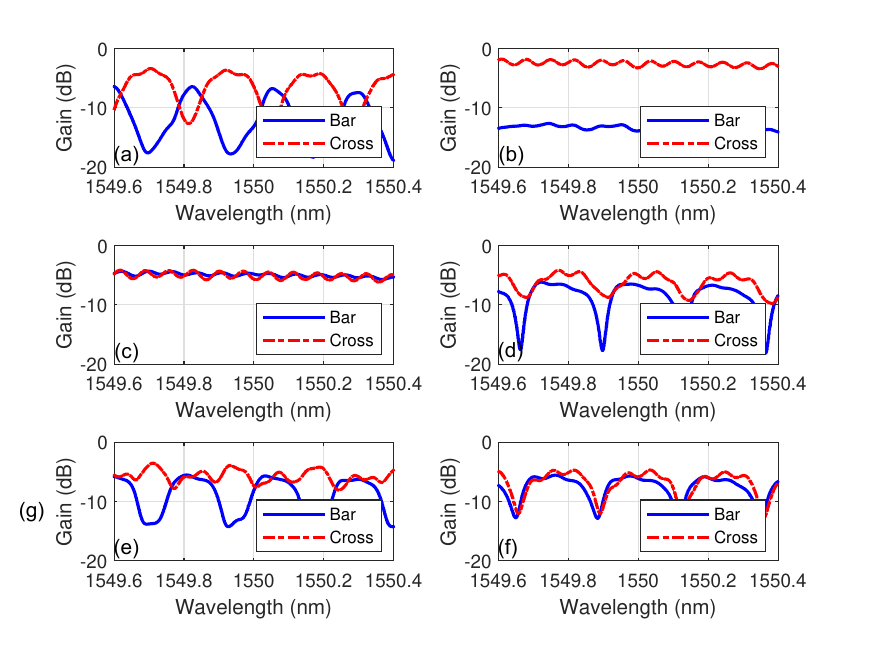}
\caption{Experimentally measured bar and cross port responses of RAMZM PIC when (a) No bias is applied, (b) Upper and lower couplers are set to zero coupling, (c) Quadrature bias is set, (d) Upper and lower couplers are set to the desired coupling coefficient (e) Upper ring anti-resonance is aligned to 1550nm, (f) Anti-resonance of both rings is aligned to 1550nm.}
\label{fig:ramzmTuningSteps}
\end{figure}

Using the steps stated in Algorithm \ref{alg:ramzmMain_algorithm}, the RAMZM was reconfigured at $\{\lambda_c, \phi_{quad}, \theta_{DC}, \tau \}=\{1550 nm, \frac{\pi}{2},\pi,\frac{1}{2}\}$. The electronic hardware backend facilitated the algorithm implementation and the intermediate steps of reconfiguration are recorded. Fig. \ref{fig:ramzmTuningSteps} shows the modulator spectral response at different stages of tuning where fiber coupling loss has been de-embedded from the spectral response. The reconfiguration process took $\sim$370s to complete.

\subsection{Highly Linear Region: $\{\phi_{quad}, \theta_{DC}, \tau \}=\{\frac{\pi}{2},\pi,\frac{1}{2}\}$}
After the reconfiguration of the modulator at $\{\lambda_c, \phi_{quad}, \theta_{DC}, \tau \}=\{1550 nm, \frac{\pi}{2},\pi,\frac{1}{2}\}$, a power sweep of two RF tones, centered around 1.1 GHz, were performed on the Device Under Test (DUT). That means the two RF tones ($f_1, f_2$) and the intermodulation (IM3) products ($2f_1-f_2, 2f_2-f_1$) were at 1.095, 1.105, 1.085 and 1.115 GHz, respectively. During this measurement, the detector current was approximately 11.23 mA, resulting in a shot-noise limited noise floor of -160 dBm/Hz.

\begin{figure}[ht!]
\centering
(a)
\includegraphics[width=1\linewidth]{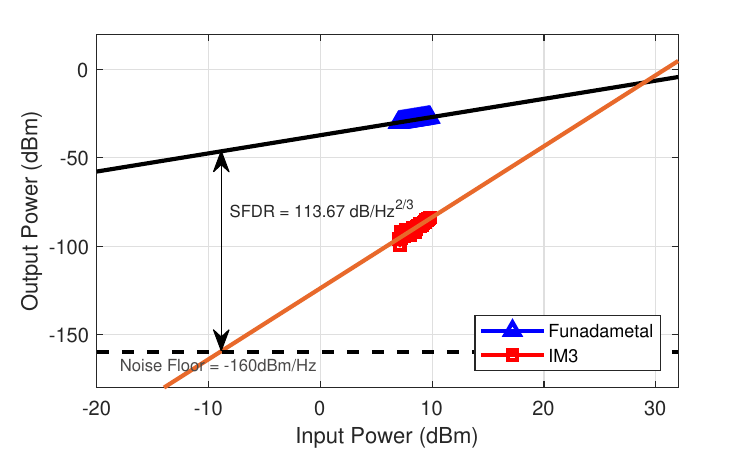}
(b)
\includegraphics[width=1\linewidth]{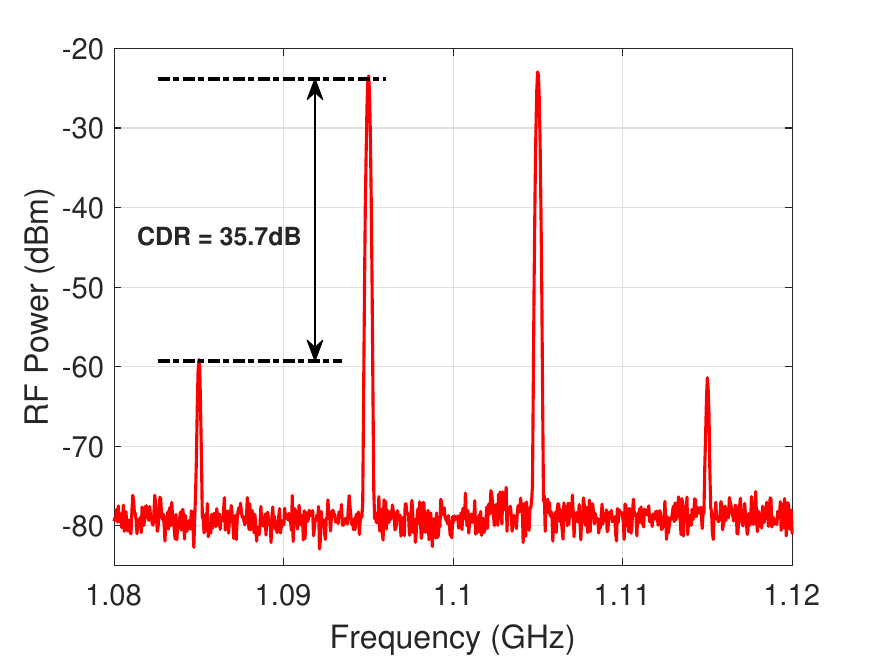}
(c)
\includegraphics[width=1\linewidth]{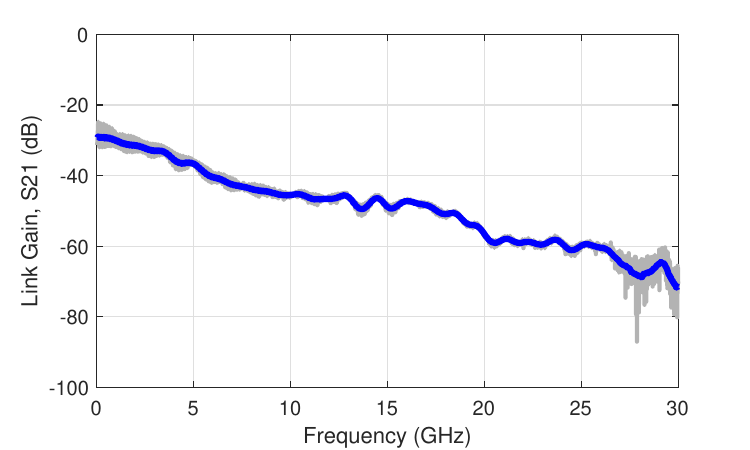}
\caption{ (a) Power of the fundamental tone and IM3 product as a function of the input power. The noise floor of -160dBm/Hz resulted in an SFDR of 113.67 $dB.Hz^{2/3}$. (b) Intermodulation spectrum of the modulator centered around 1.1 GHz. (c) EO$S_{21}$ of the modulator when input RF power is 0 dBm. For all plots, the biasing conditions are kept as $\{\lambda_c, \phi_{quad}, \theta_{DC}, \tau \}=\{1550 nm, \frac{\pi}{2},\pi,\frac{1}{2}\}$.}
\label{fig:AntiResSFDRIMSparam}
\end{figure}

The obtained results are graphically represented in Figure \ref{fig:AntiResSFDRIMSparam}, which depicts the measured SFDR, intermodulation spectrum, and S-parameters. Notably, the achieved SFDR was 113.67 $dB.Hz^{2/3}$. Additionally, the Carrier-to-Distortion Ratio (CDR) was determined to be around 35.7 dB, when the fundamental tone output power was -23.46 dBm. The S-parameter (EO$S_{21}$) plot revealed that the modulator exhibited a 3-dB bandwidth of approximately 2.5GHz.

\begin{figure}[ht!]
\centering 
\includegraphics[width=1\linewidth]{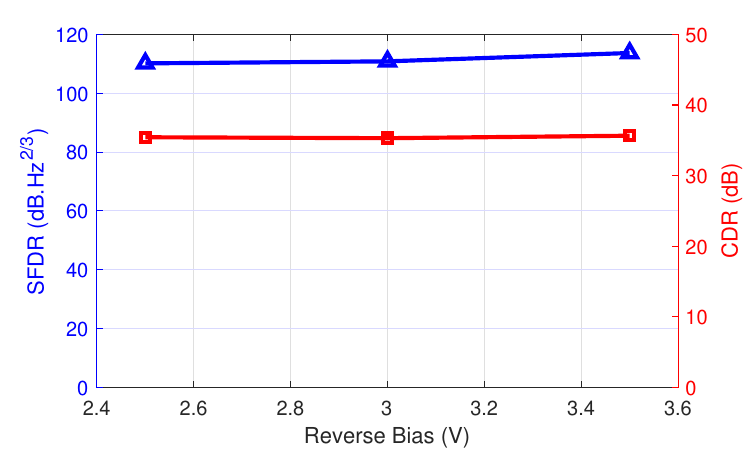}
\caption{SFDR and CDR of the modulator at different reverse biases of the PN phase modulator. The other biasing conditions are kept as $\{\lambda_c, \phi_{quad}, \theta_{DC}, \tau \}=\{1550 nm, \frac{\pi}{2},\pi,\frac{1}{2}\}$.}
\label{fig:revBiasSFDR}
\end{figure}

It is important to note that the linearity of the modulator is influenced not only by the transfer function of the RAMZM but also by the non-linearity arising from the non-linear relationship between the reverse bias voltage and the effective index change ($\Delta n_{eff}$) of the phase modulator. To investigate the impact of the reverse bias voltage on linearity, SFDR and CDR were calculated at different reverse bias voltages as shown in Fig. \ref{fig:revBiasSFDR}. Surprisingly, no significant dependence was observed, suggesting that the linearity of the phase modulator remained constant within the range of reverse bias voltages studied. This behavior could be attributed to the combined effects of plasma dispersion and the DC Kerr effect \cite{bottenfield2019silicon}.

\subsection{High Gain Region: $\{\phi_{quad}, \theta_{DC}, \tau \}=\{\frac{\pi}{2},0,\frac{1}{x}\}$}
\begin{figure}[ht!]
\centering
(a)
\includegraphics[width=1\linewidth]{figures/AntiResSFDR}
(b)
\includegraphics[width=1\linewidth]{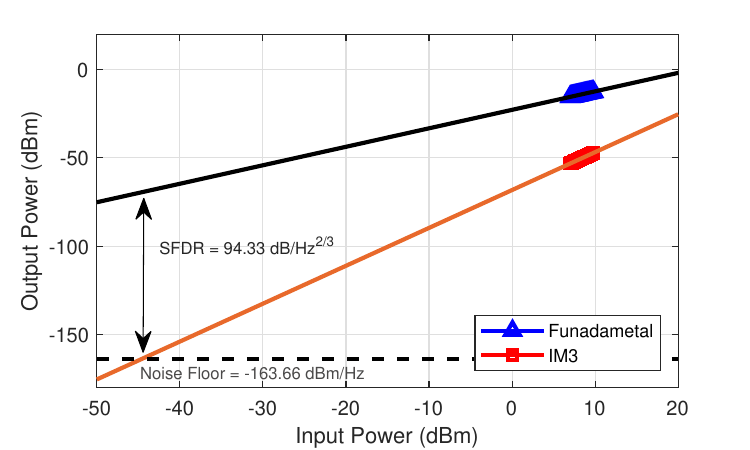}
\caption{Power of the fundamental tone and IM3 product as a function of the input power of the modulator when its at (a) Anti-resonance i.e. $\{\lambda_c, \phi_{quad}, \theta_{DC}, \tau \}=\{1550 nm, \frac{\pi}{2},\pi,\frac{1}{2}\}$. (b) Resonance i.e. $\{\lambda_c, \phi_{quad}, \theta_{DC}, \tau \}=\{1550 nm, \frac{\pi}{2},0,\frac{1}{2}\}$. The SFDR of the modulator at anti-resonance and resonance are 113.67 and 94.33 $dB.Hz^{2/3}$, respectively.}
\label{fig:ResVsAntiResSFDR}
\end{figure}

\begin{figure}[ht!]
\centering 
(a)
\includegraphics[width=1\linewidth]{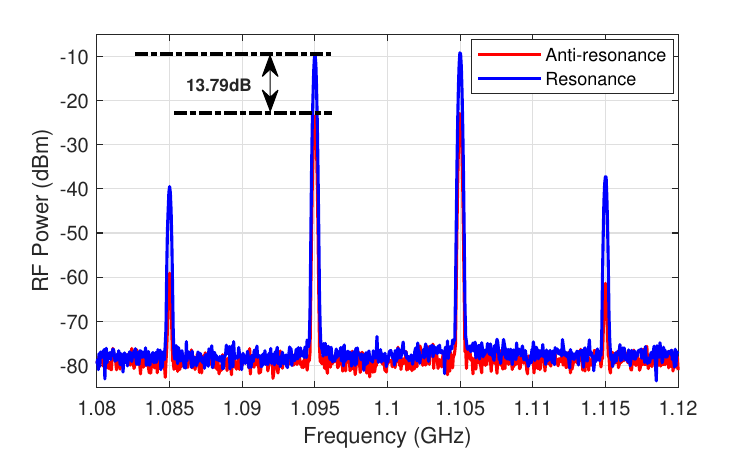}
(b)
\includegraphics[width=1\linewidth]{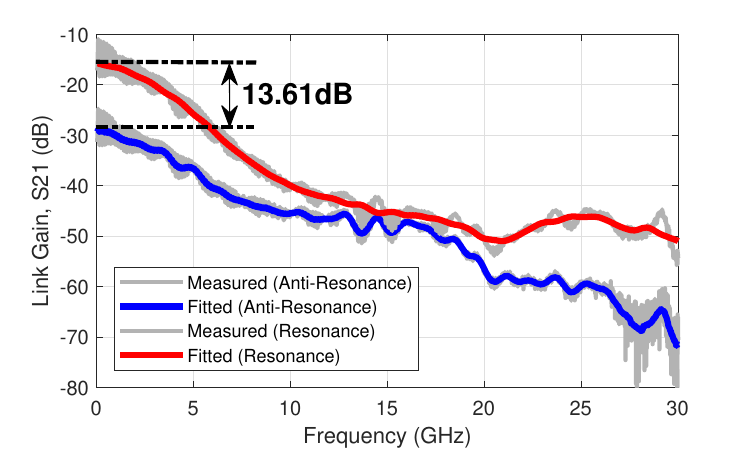}
\caption{ Comparison of (a) intermodulation spectrum of the modulator at anti-resonance i.e. $\{\lambda_c, \phi_{quad}, \theta_{DC}, \tau \}=\{1550 nm, \frac{\pi}{2},\pi,\frac{1}{2}\}$. and at Resonance i.e. $\{\lambda_c, \phi_{quad}, \theta_{DC}, \tau \}=\{1550 nm, \frac{\pi}{2},0,\frac{1}{2}\}$. (b) EO$S_{21}$ of the modulator at anti-resonance and resonance. Both (a) and (b) show more than 13 dB improvement in fundamental tone power in RAMZM biased at resonance.}
\label{fig:resVsAntiresIMSparam}
\end{figure}

 As theorized before, the RAMZM has a set of bias conditions that results in high gain at the cost of linearity (GE-RAMZM). To verify that, the modulator has been reconfigured at $\{\lambda_c, \phi_{quad}, \theta_{DC}, \tau \}=\{1550 nm, \frac{\pi}{2},0,\frac{1}{2}\}$ and two tone test was performed. As expected, the modulator linearity degraded significantly as shown in Fig. \ref{fig:ResVsAntiResSFDR}. The SFDR was found to be 94.33 $dB.Hz^{2/3}$ or about 19 dB reduction in dynamic range (DR) when compared with the modulator biased at $\{\lambda_c, \phi_{quad}, \theta_{DC}, \tau \}=\{1550 nm, \frac{\pi}{2},\pi,\frac{1}{2}\}$. The gain of the modulator, on the other hand, enhanced significantly by more than 13 dB as verified by both intermodulation spectrum and S-parameter (EO$S_{21}$) plot (shown in Fig. \ref{fig:resVsAntiresIMSparam}).

In Table \ref{tab:performance_comparison_RAMZM}, the comparison between our RAMZM and other relevant studies is presented. Here, we can see that the RAMZM provides excellent optical domain linearization, while also being compatible with CMOS foundries.

\begin{table*}[t]
\tiny
\centering
\caption{Comparison with state-of-the-art in Modulator Linearization.}
\begin{tabular}{|c|c|c|c|c|c|c|c|c|c|c|c|c|c|}
\hline
\textbf{Metric}                                                   & \cite{cardenas2013linearized
}                                         & \cite{nikolov2013advancements
}                                                                                        & \cite{jiang2015linearization
}                                               & \cite{Zhang:16
}                                              & \cite{ding2016method
}                                            & \cite{8701634
}                                                                   & \cite{bottenfield2019silicon
} & \cite{8863141
} & \cite{chen2020high
}                                           & \cite{9290115
}                                           & \cite{luo2022power
}                                                     & \cite{feng2022highly
}                                           & \begin{tabular}[c]{@{}c@{}}This\\ Work\end{tabular} \\ \hline
\begin{tabular}[c]{@{}c@{}}Material\\ Platform\end{tabular}       & Silicon                                               & LiNbO3                                                                                               & LiNbO3                                                      & \begin{tabular}[c]{@{}c@{}}III-V on\\ Silicon\end{tabular} & Silicon                                                  & Silicon                                                                         & Silicon       & Si-SiGe       & Silicon                                                 & Si-SiGe                                                 & CMOS                                                              & LiNbO3                                                  & Silicon                                                      \\ \hline
\begin{tabular}[c]{@{}c@{}}Linearization\\ Domain\end{tabular}    & Optical                                               & Optical                                                                                              & Optical                                                     & Optical                                                    & Optical                                                  & \begin{tabular}[c]{@{}c@{}}Electrical \\ + Optical\end{tabular}                 & Optical       & Electrical    & Optical                                                 & Electrical                                              & Electrical                                                        & Optical                                                 & Optical                                                      \\ \hline
Integration                                                       & Integrated                                            & Discrete                                                                                             & Discrete                                                    & Integrated                                                 & Integrated                                               & \begin{tabular}[c]{@{}c@{}}Integrated \\ + Discrete**\end{tabular}              & Integrated    & Integrated    & Integrated                                              & Integrated                                              & \begin{tabular}[c]{@{}c@{}}Integrated \\ + Discrete*\end{tabular} & Integrated                                              & Integrated                                                   \\ \hline
\begin{tabular}[c]{@{}c@{}}Foundry\\ Process\end{tabular}         & No                                                    & No                                                                                                   & No                                                          & No                                                         & -                                                        & Yes                                                                             & Yes           & Yes           & No                                                      & Yes                                                     & Yes                                                               & No                                                      & Yes                                                          \\ \hline
Technique                                                         & \begin{tabular}[c]{@{}c@{}}Dual-\\ RAMZI\end{tabular} & \begin{tabular}[c]{@{}c@{}}High Dynamic\\ Range Reciever\\ + \\ Ultra Low\\ Noise Laser\end{tabular} & \begin{tabular}[c]{@{}c@{}}Dual Parallel\\ MZM\end{tabular} & \begin{tabular}[c]{@{}c@{}}Dual-\\ RAMZI\end{tabular}      & \begin{tabular}[c]{@{}c@{}}Doping\\ Control\end{tabular} & \begin{tabular}[c]{@{}c@{}}DC Kerr \\ Ring \\ + \\ IM3\\ Injection\end{tabular} & DC Kerr       & Co-design     & \begin{tabular}[c]{@{}c@{}}Single-\\ RAMZI\end{tabular} & \begin{tabular}[c]{@{}c@{}}IM3\\ Injection\end{tabular} & Predistortion                                                     & \begin{tabular}[c]{@{}c@{}}Single-\\ RAMZI\end{tabular} & \begin{tabular}[c]{@{}c@{}}Dual-\\ RAMZI\end{tabular}        \\ \hline
\begin{tabular}[c]{@{}c@{}}SFDR \\ Frequency\\ (GHz)\end{tabular} & 1-10                                                  & 1-18                                                                                                 & 11.95                                                       & 10                                                         & 2-20                                                     & 1.2                                                                             & 1             & 1-20          & 1                                                       & 0.5-20                                                  & 20-35                                                             & 1                                                       & 1.1                                                          \\ \hline
\begin{tabular}[c]{@{}c@{}}SFDR\\ (dB.Hz2/3)\end{tabular}         & 99-106                                                & 109-114                                                                                              & 116                                                         & 117                                                        & 110-113.7                                             & 98-108                                                                          & 110           & 101-109       & 111.3                                                   & 109-120                                                 & 99-101                                                            & 102.31                                                  & 113                                                          \\ \hline
Gain                                                              & -22                                                   & -4/-15                                                                                               & -24                                                         & -17/-22                                                    & -24                                                      & -31                                                                             & -19/-35       & -1/-14        & -24/-31                                                 & -                                                       & -                                                                 & -44                                                     & -36.8                                                          \\ \hline
Bias Control                                                      & Manual                                                & Manual                                                                                               & Manual                                                      & Manual                                                     & Manual                                                   & Manual                                                                          & Manual        & Manual        & Manual                                                  & Manual                                                  & Manual                                                            & Manual                                                  & Automatic                                                    \\ \hline
\end{tabular}
  \label{tab:performance_comparison_RAMZM}
\end{table*}

\section{Conclusion}
In conclusion, We presented a detailed analysis of silicon photonic RAMZM-based analog optical link that provides significant improvement in linearity over the MZM-based links. A RAMZM PIC that can be reconfigured at different biasing regimes based on the user specs has been demonstrated. Leveraging the tuning algorithm, we experimentally demonstrated the optical domain linearization of RAMZM with SFDR $>$ 113dB.Hz$^{2/3}$ at 1.1GHz. Moreover, the presented RAMZM PIC is fabricated in a CMOS-compatible SiP foundry process. By taking full advantage of the low-cost CMOS fabrication process, this reconfigurable EO modulator can be widely deployed as a plug-n-play module in integrated RF photonic SoCs.

\section{Acknowledgment}
The authors gratefully acknowledge the generous funding support from Air Force Office of Sponsored Research (AFOSR) YIP Award FA9550-17-1-0076, DARPA YFA Award HR00112110001, NSF CAREER Award EECS-2014109.



%


\ifCLASSOPTIONcaptionsoff
  \newpage
\fi


\bibliographystyle{IEEEtran}
\bibliography{RAMZM_LINK}

\end{document}